\documentclass[lettersize,journal]{IEEEtran}
\usepackage{amsmath,amsfonts}
\usepackage{algorithmic}
\usepackage{array}
\usepackage[caption=false,font=normalsize,labelfont=sf,textfont=sf]{subfig}
\usepackage{textcomp}
\usepackage{cite}
\usepackage{stfloats}
\usepackage{url}
\usepackage[normalem]{ulem}
\usepackage{verbatim}
\usepackage{graphicx}
\usepackage{gensymb}
\usepackage{amsmath,amssymb,amsfonts}
\usepackage{dsfont}
\usepackage[dvipsnames]{xcolor}
\usepackage{comment} 

\usepackage{balance}

\newcommand{\ex}{{\rm e}}

\begin{document}
\title{Phase Noise Detection via Expectation Propagation and Related Algorithms 
        \thanks{This research was granted by University of Parma through the action Bando di Ateneo 2022 per la ricerca co-funded by MUR-Italian Ministry of Universities and Research - D.M. 737/2021 - PNR - PNRR - NextGenerationEU.}
}
\author{Elisa Conti, Armando Vannucci, Amina Piemontese, and Giulio Colavolpe}

%\markboth{IEEE Transactions on Communications,~Vol.~x, No.~x, Month~Year}%
%{\title}

\maketitle

\begin{abstract}
In the context of signal detection in the presence of an unknown time-varying channel parameter, receivers based on the Expectation Propagation (EP) framework appear to be very promising.  EP is a message-passing algorithm based on factor graphs with an inherent ability to combine prior knowledge of system variables with channel observations. This suggests that an effective estimation of random channel parameters can be achieved even with a very limited number of pilot symbols, thus increasing the payload efficiency. However, achieving satisfactory performance often requires ad-hoc adjustments in the way the probability distributions of latent variables - both data and channel parameters - are combined and projected.
Here, we apply EP to a classical problem of coded transmission on a strong Wiener phase noise channel, employing soft-input soft-output decoding. We identify its limitations and propose new strategies which reach the performance benchmark while maintaining low complexity, with a primary focus on challenging scenarios where the state-of-the-art algorithms fail.
\end{abstract}


\section{Introduction}
\IEEEPARstart{D}{etection} algorithms for channels affected by a time-varying phase noise have received a lot
of attention in the literature considering either linear or continuous-phase modulations
and different scenarios (see, e.g., \cite{CBC_JSAC_2005, BaCo07b, Co14, PeShGa00, KrKhErSv13, KreimerRaphaeliTCOM2018, MeNaBlEr12} and references therein).
This is because in many communication links phase noise must be considered one of the major impairments. 
Examples are represented by scenarios employing a high carrier frequency, such as (coherent) optical communications, communications from geostationary satellites, etc. 

One of the most effective algorithms is that proposed in~\cite{CBC_JSAC_2005}. It is designed using the framework based on factor graphs~(FGs) and the sum-product algorithm~(SPA). In a scenario like the one at hand, where  continuous random variables (the channel phase) appear in the FG, a common approach to implement the SPA is to resort to the use of canonical distributions~\cite{WoSt94}. In particular, in~\cite{CBC_JSAC_2005}, the messages representing the a-posteriori probability density functions~(pdfs) of the channel phase are represented through Tikhonov distributions, which can be described by a single complex parameter. 
Although suboptimal, the resulting algorithm exhibits an excellent trade-off between performance and complexity. The suboptimality is related to the joint presence of discrete (the code symbols) and continuous random variables in the graph, which brings up mixture pdfs with exponential proliferation, approximated in~\cite{CBC_JSAC_2005} with unimodal distributions. The presence of distributed pilot symbols is thus required to make the algorithm bootstrap.
In fact, the algorithm in~\cite{CBC_JSAC_2005} based on the Tikhonov parametrization is very sensitive to the positioning of pilot symbols, so that increasing the pilot spacing causes a performance degradation. 
In particular, when pilot symbols are concentrated at both ends of the codeword, as a preamble/postamble, or, more in general, when a maximum distance between consecutive pilot blocks is exceeded, it is known to fail~\cite{CBC_JSAC_2005,ColavolpeModenini2013}. 
This occurs since this kind of algorithm cannot achieve any effective phase detection without the aid of extrinsic information, provided by the decoding part of the FG during turbo iterations, or without distributed pilots (with sufficiently close spacing). 
Therefore, other alternative solutions have to be analyzed to this purpose. 

As an extension of~\cite{CBC_JSAC_2005}, the results of Raphaeli and coworkers~\cite{ShayovitzRaphaeliTCOM2016,KreimerRaphaeliTCOM2018} are obtained by letting mixture messages propagate one step further into the FG; their exponential proliferation being limited by an appropriate pruning of the mixture components, performed at the level of the Markov chain governing phase noise. 
This way, the phase uncertainty inherent in the observation of channel output is free to interact with the provisional estimation of previous phase samples (or with the following ones, in backward block processing). 
Such a mixture message reduction approach, 
however, 
is characterized by a considerable complexity.

The same objective of letting channel observations interact with provisional estimates of adjacent phase samples is achieved by the expectation propagation~(EP) algorithm~\mbox{\cite{Minka_UAI_2001, Minka_TechRep_2005, HeskesZoeter_UAI_2002, Gelman_arXiv_2014}}, where similar mixture reduction techniques are applied to the marginal distributions of the variables of interest, rather than to individual messages. 
Despite the projection of messages or marginals do have some features in common, and can even reduce to the same algorithm in some cases~\cite{VannucciColavolpeVeltri_CommLett_2020_ImpNoise, MirbadinVannucciColavolpe_TSP_2020, MirbadinVannucciColavolpePecoriVeltri_ApplSci_2021}, there is a profound conceptual difference between them. 
In fact, only in the latter case, a message coming from a channel observation is merged, i.e., multiplied, with the (provisional) {\em prior belief} on the destination variable that comes from the rest of the FG. 
This is the key to the potential success of EP in many similar applications, including, e.g., transmission over fading channels~\cite{QiMinka_TrWC_2007, CoPiCoVa23}. 

The EP framework has already been successfully applied to phase noise channels, with either distributed or concentrated pilots~\cite{ColavolpeModenini2013,SzczecinskiBouaziziAhikam2020arXiv}, resorting to joint detection and decoding. 
This is not, however, a practical solution, since a separate (and sequential) detection of channel parameters and decoding is desirable in the design of digital receivers. On the contrary, the challenge of making EP work with separate phase detection and decoding has not been solved.
In fact, previous analyses have shown that, despite its ability to bootstrap the phase estimation process from a proper block of preamble/postamble pilots, the native EP algorithm is not able to effectively refine the phase estimates beyond the first phase-detecting iteration~\cite{CoCoPiVa23}; and this is true even when pilot blocks are close enough, a situation where the algorithm in~\cite{CBC_JSAC_2005} shows a better performance.  

In this work, we first examine the EP framework applied to phase noise channels, identify its weaknesses and the underlying reasons. Preliminary results of this analysis can be found in~\cite{CoCoPiVa23}. We then propose novel strategies to improve its performance, resulting in a flexible algorithm capable of adapting to various pilot positioning and iterations scheme (whether separate or joint detection and decoding) always outperforming the state-of-the-art solutions.

The rest of this paper is organized as follows. In Section~\ref{sec:System_FG}, we introduce the system model and the FG representation of the joint \textit{posterior} pdf of the information symbols and channel parameters. Section~\ref{sec:EP} is devoted to the description of the EP algorithm while, in Section~\ref{sec:ModEP}, we present the proposed strategies to overcome the EP limitations without increasing its complexity. The message scheduling is outlined in Section~\ref{sec:sched}. A complexity comparison among the analyzed algorithms is carried out in Section~\ref{sec:Complexity}. Finally, in Section~\ref{sec:Results} we show and discuss the numerical results and, in Section~\ref{sec:Conclusions}, we draw some concluding remarks.

\textit{Notation.} The following notational convention will be used throughout the paper. Bold letters denote vectors, the superscript $(\cdot)^*$ applied to a scalar term denotes its complex conjugate. We indicate with $\Re[\cdot]$ and $\Im[\cdot]$ the real and imaginary parts of a complex quantity, respectively. A complex circularly symmetric (respectively, real) Gaussian random variable ${x}$ with mean $\eta_{x}$ and variance $\sigma_{x}^2$ is denoted by \mbox{$x \sim {\cal N}_{\mathbb C}(\eta_x,\sigma^2_x)$} [respectively, by \mbox{$x \sim {\cal N}(\eta_x,\sigma^2_x)$}]. We denote the complex (respectively, real) Gaussian pdf with argument $x$, mean $\eta_{x}$, variance $\sigma_{x}^2$ by \mbox{$g_{\mathbb C}({x}-\eta_{x}; \sigma_{x}^2 )$} [respectively, by \mbox{$g({x}-\eta_{x}; \sigma_{x}^2 )$}]. We denote by $p(\cdot )$ the pdf of continuous variables (vectors) and with a capital $P(\cdot)$ the probability mass function (pmf) of discrete variables as well as the pdf of mixed random vectors. We use $E_p[\cdot ]$  to indicate the expectation under the distribution $p(\cdot)$. When $p(\cdot)$ is not specified, it is implicit from the context.

\section{System Model and Related Factor Graph}\label{sec:System_FG}

\begin{figure}
\begin{centering}
\includegraphics[width=0.95\columnwidth, clip=true, trim = 15mm 0mm 15mm 0mm]{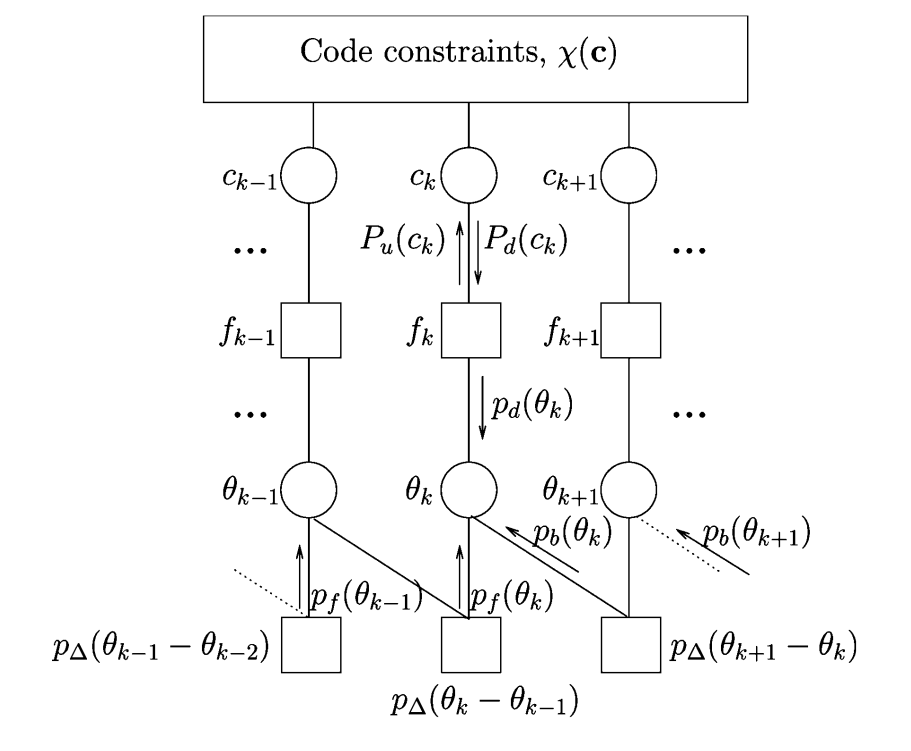}
% order of 'trim = left bottom right top
\par\end{centering}
\caption{Factor graph representing the joint probability distribution of phase noise samples and coded symbols.\label{fig:FG_CBC}}
\end{figure}

In the system we wish to investigate, a sequence of $K$ complex coded symbols $\boldsymbol{c}=[c_0,c_1,\cdots,c_{K-1}]$, modulated over an $M$-ary constellation (with alphabet $\mathcal{A}$), is transmitted over an additive white Gaussian noise (AWGN) channel also affected by Wiener phase noise, so that the received samples are
\begin{equation} \label{eq:rk}
r_k= c_k \ex^{j\theta_k} + n_k \, ,
\end{equation}
where $\{ n_k\}$ is a sequence of independent (complex and circularly symmetric) Gaussian noise samples, i.e., \mbox{$n_k \sim {\cal N}_{\mathbb C}(0,2\sigma^2)$}, with zero mean and given variance per component $\sigma^2$, while the phase noise sequence \mbox{$\boldsymbol{\theta}=[\theta_0,\theta_1,\cdots ,\theta_{K-1}]$} follows the Wiener model, where each sample
\begin{equation} \label{eq:thetak}
\theta_k= \theta_{k-1} + \Delta_k
\end{equation}
results from the previous one plus a zero-mean (real) Gaussian increment $\Delta_k \sim {\cal N}(0,\sigma_{\Delta}^2)$ whose variance $\sigma_{\Delta}^2$ dictates the severity of phase noise.

Since the system model is the same as the one discussed in~\cite{CBC_JSAC_2005}, the FG representing the joint distribution of the latent variables $\boldsymbol{\theta}$ and $\boldsymbol{c}$, conditioned on the value of the observed variables $\boldsymbol{r}=[r_0,r_1,\cdots ,r_{K-1}]$, is the same as the one shown therein and is reported in Fig.~\ref{fig:FG_CBC}. The subscripts of messages identify their direction in the FG ($u$p, $d$own, $f$orward, $b$ackward), whereas the subscript of variable and factor nodes coincides with a time index.

Before observing the channel output, the phase noise sequence is independent of that of coded symbols, so that the joint posterior distribution of all latent system variables is 
\begin{equation} \label{eq:Pcth_r}
\begin{split}
 P(\boldsymbol{c,\theta \!\! \mid \! r}) 
&\!\propto P(\boldsymbol{c})p(\boldsymbol{\theta})p(\boldsymbol{r \mid c,\theta}) \\
  & \propto \chi(\boldsymbol{c}) p(\theta_0) \prod_{k=1}^{K-1} p(\theta_k \mid \theta_{k-1}) \prod_{k=0}^{K-1} f_k(c_k, \theta_k)
\end{split}
\end{equation}
where $\propto$ is the proportionality symbol, $\chi(\boldsymbol{c})$ is an indicator function implementing the code constraints and we assume that all the allowed codewords in the employed codebook are equally likely. 

The factor node $f_k$ implements the observation of the received samples in~(\ref{eq:rk}), hence 
\begin{equation} \label{eq:fk_Gauss}
\begin{split}
f_k(c_k, \theta_k) & = p(r_k \mid c_k, \theta_k) = g_{\mathbb C}( r_k - c_k \ex^{j\theta_k}; \sigma^2 ) \\
& \propto \exp\left( -\frac{1}{2\sigma^2} \left| r_k - c_k \ex^{j\theta_k} \right|^2 \right) 
\end{split}
\end{equation}
is a complex Gaussian distribution. 

Besides $f_k$, the only other factor nodes in the FG of Fig.~\ref{fig:FG_CBC} (if we disregard the check nodes included in the code sub-graph) are the ones related to the conditional pdfs in~(\ref{eq:Pcth_r}) that implement the Markov chain governing the Wiener phase noise, 
\begin{equation} \label{eq:pDelta}
p(\theta_k \mid \theta_{k-1}) \triangleq p_{\Delta}(\theta_k - \theta_{k-1}) = g( \theta_k - \theta_{k-1}; \sigma_{\Delta}^2 )\,,
\end{equation}
which are Gaussian too, as per (\ref{eq:thetak}). 

Owing to the fact that the additive sources of randomness in~(\ref{eq:rk}) and~(\ref{eq:thetak}) are Gaussian, the FG in Fig.~\ref{fig:FG_CBC} seems to represent a Gaussian belief network, where all messages are Gaussian~\cite{Seeger_PhD_2003} and the message passing procedure resembles the simple operations of a Kalman smoother. 
% NO: solo qua...  % \cite{BarberCemgil_SPmag_2010}
This is clearly not true, for two reasons. 
First, the problem at hand entails a ``mixed model", where continuous~($\boldsymbol{\theta}$) and discrete~($\boldsymbol{c}$) variables coexist, which produces probabilistic mixtures in the FG, specifically arising from the  messages $p_d(\theta_k)$ which are linear combinations of simpler  component distributions. 
In addition, (\ref{eq:fk_Gauss}) is a Gaussian pdf only if seen as a function of~$r_k$ whereas the factor nodes~$f_k$ send downward messages~$p_d(\theta_k)$ to the variable nodes~$\theta_k$, hence~(\ref{eq:fk_Gauss}) must be seen as a function of~$\theta_k$ and is thus  proportional to a Tikhonov distribution~\cite{CBC_JSAC_2005}, 
\begin{equation} \label{eq:fk_Tikh}
\begin{split}
f_k(c_k, \theta_k) & \propto \exp\left( -\frac{|c_k|^2}{2\sigma^2} + \frac{1}{\sigma^2}\Re\left[ r_k c_k^* \ex^{-j\theta_k} \right] \right) \\
& = \exp\left( -\frac{|c_k|^2}{2\sigma^2} \right)2\pi I_0(|z_k |) t(\theta_k ; z_k)
\end{split}
\end{equation}
where $z_k = r_k c_k^* /\sigma^2$ is the complex parameter characterizing the general Tikhonov distribution:
\begin{equation} \label{eq:Tikhonov}
% \begin{split}
t(\theta_k ; z_k) = \frac{1}{2\pi I_0(|z_k |)} \exp\left( \Re\left[ z_k \ex^{-j\theta_k} \right] \right) \,.
% \end{split}
\end{equation}
In directional statistics~\cite{mardia2000directional}, the phase $\angle z_k$ corresponds to the {\em circular mean} of the circular random variable~$\theta_k$ in~(\ref{eq:Tikhonov}), while the magnitude $|z_k|$ is a measure of its precision, i.e., of the inverse of the variance $\sigma^2_{\theta_k}=1-\frac{I_1(|z_k|)}{I_0(|z_k|)} \in [0;1]$, where $I_p(x)$ is the modified Bessel function of the first kind of order $p$.\footnote{More in general, the $p$-th trigonometric moment of the Tikhonov variable~$\theta_k$ is $E_t[\exp{(jp\theta_k)}]=\exp{(jp\angle{z_k})}\frac{I_p(|z_k|)}{I_0(|z_k|)}$, so that the phase of the first trigonometric moment is the circular mean.}

The Bayesian inference problem for the estimation of symbols and channel parameters cannot be solved by the plain SPA since the message 
\begin{equation}\label{eq:pd_mix}
p_d(\theta_k) = \sum_{m=0}^{M-1} P_d (c_k^m) f_k(c_k^m, \theta_k) \propto \sum_{m=0}^{M-1} \alpha_k^m t(\theta_k ; z_k^m) 
\end{equation}
is a Tikhonov mixture. Its coefficients $\alpha_k^m$, defined as
\begin{equation}\label{eq:alpha_m}
    \alpha_k^m = P_d (c_k^m) \exp\left( -\frac{|c_k^m|^2}{2\sigma^2} \right)2\pi I_0(|z_k^m |)\,,
\end{equation}
depend on the extrinsic information on each symbol (i.e., on the values of its pmf for each of the possible $M$ symbols $c_k^m$), as provided by the soft-input soft-output decoding part of the FG (top of Fig.~\ref{fig:FG_CBC}), as well as on the magnitude of the symbols. 
% (unless one employs constant magnitude constellations, such as M-PSK, in which case $|c_k^m|$ are constant, at every time epoch). 
In the SPA, the mixture messages~(\ref{eq:pd_mix}) would propagate through the bottom half of the FG and eventually proliferate to produce untractable mixtures with an exponentially increasing number of components. 
Approximate inference is thus demanded, which can be performed in a variety of ways.

%%%%%%%%%%%%%%%%%%%%%%%%%%%%%%%%%%%%%%%%%%%%%%%%%%%%%%%%%
\section{Transparent Propagation Algorithms and Expectation Propagation}\label{sec:EP}

The most straightforward approach to approximate message passing is to discretize the distribution of~$\theta_k$, with a given number of samples $N_{\theta}$, so that all the latent system variables appear to be discrete and modelled by their pmf. 
This produces an algorithm called discretized-phase BCJR~\mbox{(dp-BCJR)} in~\cite{PeShGa00}, since the forward-backward message passing procedure, along the Markov chain in the bottom half of the FG, is the same as that of the celebrated BCJR algorithm.  
Its computational complexity is very high, thus we shall implement this algorithm, denoted as {\em dp-BCJR} in the simulation results, as a practical benchmark, assuming that its performance gets close to that of an ideal SPA when $N_{\theta}$ is sufficiently large.

A totally different approach relies on projecting messages and/or marginals onto a selected family of approximating distributions, usually in the exponential form (so that the family is closed under the multiplication operation), \begin{equation}
q(x)=\exp\left(\sum_i \eta_i g_i(x)\right)\,,
\end{equation}
where $\eta_i$ are called natural parameters and the functions $g_i(x)$ are the features of the family. 
This way, only the natural parameters $\eta_i$ are updated, for each distribution, resulting in a parametric message passing procedure. 
In \cite{CBC_JSAC_2005}, the Tikhonov approximating family is selected quite naturally, since it is known to be the marginal distribution of $\theta_k$, conditioned on the values taken by the corresponding symbol and by the corresponding channel output, i.e., $p(\theta_k \mid c_k, r_k) \propto f_k(c_k, \theta_k)$, as per (\ref{eq:fk_Gauss}). 
Curiously, the way in which message $p_d(\theta_k)$ has been projected onto a Tikhonov pdf in~\cite{CBC_JSAC_2005} relies on the Gaussian expression~(\ref{eq:fk_Gauss}) of the factor node, so that~(\ref{eq:pd_mix}) is approximated by the Gaussian pdf with minimum Kullback-Leibler (KL) divergence from the mixture, which is further interpreted as a Tikhonov message towards~$\theta_k$.

A more natural solution would have been to project the Tikhonov mixture~(\ref{eq:pd_mix}) onto a Tikhonov pdf with minimum KL divergence, which is achieved (for this as well as for any other exponential approximating family) by matching the expectations of the features~\cite{Minka_TechRep_2005}. We refer to this approach, along with the Tikhonov parameterization of~\cite{CBC_JSAC_2005}, as Transparent Propagation~(TP) algorithms (the reason will be clear later). 
The features of a Tikhonov pdf like~(\ref{eq:Tikhonov}) are $\cos(\theta_k )$ and $\sin(\theta_k )$, associated with the natural parameters $\Re [z_k]$ and $\Im [z_k]$ respectively, so that their expectations can be jointly computed as the $\Re / \Im$ parts of the first trigonometric moment $E[\exp (j\theta_k)]$. 
Exploiting known results~\cite{mardia2000directional} and the linearity of expectation, the mixture in~(\ref{eq:pd_mix}) is approximated by the Tikhonov pdf $p_d^{TP}(\theta_k)= t(\theta_k; z_k^{TP})$, which achieves the minimum KL divergence when $z_k^{TP}$ obeys the following {\em moment matching} equation 
\begin{equation}\label{eq:MomentMatch}
\frac{I_1(|z_k^{TP}|)}{I_0(|z_k^{TP}|)} \ex^{j\angle z_k^{TP}} = \sum_{m=0}^{M-1} \overline{\alpha}_k^m  \frac{I_1(|z_k^m|)}{I_0(|z_k^m|)} \ex^{j\angle z_k^m}
\end{equation}
where $\overline{\alpha}_k^m$ stands for the normalized version of the coefficients in~(\ref{eq:pd_mix}), i.e., 
\begin{equation}\label{eq:alpha_norm}
    \sum_{m=0}^{M-1} \overline{\alpha}_k^m =1\,.
\end{equation} 
The complex equation~(\ref{eq:MomentMatch}) yields the circular mean $\angle z_k^{TP}$ as well as the variance of phase noise, related to the Bessel ratio of the magnitude~$|z_k^{TP}|$ \cite{ShayovitzRaphaeliTCOM2016}. 
% \footnote{Such a result has been restated in the form of a theorem in \cite{ShayovitzRaphaeliTCOM2016}.}
% $A(|z_k^{TP}|)$ of (\ref{eq:MomentMatch}), hence the variance $\sigma^2_{\theta_k}=1-A(|z_k|)$ \cite{mardia2000directional}. 
Despite the different analytical approach, the two ways of projecting the mixture message $p_d(\theta_k)$ described above, i.e., that of~\cite{CBC_JSAC_2005} and the one in~(\ref{eq:MomentMatch}), yield extremely similar results, as verified by numerical simulation in all phase noise scenarios that we analyzed. For this reason, their equivalent performance is collectively shown under the {\em TP} label in the simulation results in Section~\ref{sec:Results}. 

In the EP algorithmic framework~\cite{Minka_UAI_2001}, it is the entire marginal of each latent variable that is approximated/projected; in our case,  
\begin{equation}\label{eq:def_pu}
p(\theta_k)= p_d(\theta_k)p_f(\theta_k)p_b(\theta_k) = p_d(\theta_k)p_u(\theta_k)
\end{equation}
of which the message $p_d(\theta_k)$ only represents one factor. 
In~(\ref{eq:def_pu}), message $p_u(\theta_k) \triangleq p_f(\theta_k)p_b(\theta_k)$ is defined as the product of forward and backward messages, and is sent upwards, opposite to message $p_d(\theta_k)$, although not reported in the FG in Fig.~\ref{fig:FG_CBC}.
Indeed, the conceptual meaning of the two factors of the marginal in~(\ref{eq:def_pu}) is the following: $p_d(\theta_k)$ is an {\em observation message}, that carries information on $\theta_k$ after the channel observation entailed in the factor node~(\ref{eq:fk_Tikh}), while $p_u(\theta_k)$ plays the role of a temporary {\em prior belief} for the phase noise variable $\theta_k$, as provided by the rest of the FG. 
Denoting the projection operation as\footnote{We denote with $\text{KL}[p(\cdot)\,||\,q(\cdot)]$ the KL divergence of an approximated distribution $q(\cdot)$ from the exact distribution $p(\cdot)$.} 
\begin{equation}
\textrm{proj}[p(\cdot )]=\arg\min_{q(\cdot )\in {\cal F}} \text{KL}\left[ p(\cdot )\parallel q(\cdot )\right]\,,
\end{equation}
where $\cal F$ is the approximating family, the approximating message
\begin{equation}\label{eq:pd_EP}
p_d^{EP}(\theta_k)= \frac{\textrm{proj}[p_d(\theta_k)p_u(\theta_k)]}{p_u(\theta_k)}
\end{equation}
is computed and sent in place of the mixture $p_d(\theta_k)$. 
After the observation $f_k(c_k, \theta_k)$, the estimated marginal for the phase noise variable $\theta_k$ is thus  updated in a different way, compared to the plain SPA rule (\ref{eq:def_pu}), and expressed as
\begin{equation}
    p^{EP}(\theta_k)= p_d^{EP}(\theta_k)p_u(\theta_k) \, ,
\end{equation} 
which is  equal to the numerator of (\ref{eq:pd_EP}) and thus belongs, by construction, to the approximating family (Tikhonov, in our case). 

The projection operation in~(\ref{eq:pd_EP}) still amounts to a moment matching operation like~(\ref{eq:MomentMatch}), despite a marginal is projected, in EP, rather than a message, as in TP. 
Of course, in the case of EP, the temporary prior $p_f(\theta_k)p_b(\theta_k)$ must be accounted for in the right hand side of (\ref{eq:MomentMatch}), by simply adding the (complex) parameters $z_{k,f}$ and $z_{k,b}$ of the forward and backward messages to those of each mixture component, i.e., to $z_k^m$ (hence the coefficients $\alpha_k^m$ must be calculated accordingly). 
{Therefore, from a conceptual standpoint, the mixture to be projected is now the result of the interaction between the prior belief, deriving from the Markov chain at the bottom half of the FG in Fig.~\ref{fig:FG_CBC}, and the current channel observation, represented by message $p_d(\theta_k)$. 
The expression for the marginal mixture parameters thus becomes 
\begin{equation}\label{eq:z_MIX}
    z_{k,MIX}^{m} = z_{k,u}+z^m_k
\end{equation}
where we denoted with $z_{k,u}=z_{k,f}+z_{k,b}$ the temporary prior parameter.}

The general approach to approximate message passing can be either a projection of the marginal pdf of each variable node (onto the selected approximating family), as prescribed by EP, or a projection of individual mixture messages onto the same family, as described above for TP algorithms. 
If, for the sake of argument, the EP algorithm did not produce any information about the temporary prior, then we would assume $p_f(\theta_k)p_b(\theta_k)$ to be uniform in~(\ref{eq:pd_EP}) and hence it would {\em transparently shift out} of the projection operation, to be simplified with the denominator, so that~(\ref{eq:pd_EP}) would reduce to the simple projection of message~$p_d(\theta_k)$. This is the reason for which we denote the message projection approach as TP~\cite{MirbadinVannucciColavolpe_TSP_2020}.
It can be demonstrated that in some specific problems, the TP and EP approaches lead to the same solution~\cite{MirbadinVannucciColavolpePecoriVeltri_ApplSci_2021}, although in general they  differ remarkably. 

No matter if one of the TP algorithms or if EP is adopted, the forward and backward messages $p_f(\theta_k)$ and $p_b(\theta_k)$ are assumed to belong to the Tikhonov family too, despite the factor nodes in the Markov chain of the FG, e.g., \mbox{$p(\theta_{k+1} \mid \theta_k)$}, introduce a convolution between the Tikhonov message $p_d^{TP/EP}(\theta_k)p_f(\theta_k)$ and a Gaussian like~(\ref{eq:pDelta}), for the computation of $p_f(\theta_{k+1})$. 
As it is shown in the Appendix of~\cite{CBC_JSAC_2005}, if the phase noise variance $\sigma_\Delta^2$ is not exaggeratedly large then $p_f(\theta_{k+1})$ is very well approximated by a Tikhonov pdf whose parameter $|z_{f,k+1}|$  is less than that of the incoming message, $|z_{f,k}+z_k^{TP/EP}|$, while the circular mean is the same (see~(36)-(38) in~\cite{CBC_JSAC_2005}, for details). 
This corresponds conceptually to the fact that the Wiener phase noise update in~(\ref{eq:thetak}) does not bias the temporary estimate for~$\theta_{k+1}$ but decreases its precision, according to its Gaussian variance. A similar conclusion holds for the backward messages too. 

%%%%%%%%%%%%%%%%%%%%%%%%%%%%%%%%%%%%%%%%%%%%%%%%%%%%%%%%%
\section{The EP Modification}\label{sec:ModEP}

The EP algorithm has been already applied to transmission over Wiener phase noise channels in~\cite{ColavolpeModenini2013} (and later in~\cite{SzczecinskiBouaziziAhikam2020arXiv}), in the absence of distributed pilot symbols, achieving a good performance when turbo iterations\footnote{{We call {\em turbo} iterations the exchange of information between the detector and decoder parts of the FG, implemented by the vertical messages~$p_d(\cdot)$ and $P_u(\cdot)$ in Fig.~\ref{fig:FG_CBC}.}} are employed. 
However, existing literature does not demonstrate successful operation of EP with separate phase detection and decoding.

As it is known, the performance of TP algorithms, such as~\cite{CBC_JSAC_2005}, significantly deteriorates by increasing the pilot spacing, leading to complete failure when they are concentrated at both ends of the codeword.
In contrast, in these scenarios, EP represents a viable alternative, due to its ability to self-sustain the process of phase estimation across a long block of payload coded symbols. 

However, the native EP algorithm is not able to effectively refine the phase estimates beyond the first phase-detecting iteration~\cite{CoCoPiVa23}, resulting in unsatisfying performance both with distributed pilots (where it is outperformed by TP algorithms) and with more concentrated pilots (where, unable to benefit from inner detector iterations, its performance is limited as shown in Section~\ref{sec:Results}). 
 
Through a detailed numerical analysis of EP, we found that its failure is confined to some critical data packets that, due to a long sequence of noisy observations, bring the sequence of precision values $|z^{EP}_{k,f/b}|$ to a collapse, while propagating forward or backward messages, $p_f(\theta_k )$ and $p_b(\theta_k )$. 
Once a critical lower threshold is exceeded, the precision approaches zero and is rarely able to recover, even if  the channel in (\ref{eq:rk}) outputs reliable observations, i.e., samples~$r_k$ with little noise. 
In order to make EP overcome these limiting situations, we introduce the modifications described hereafter, which improve its robustness even in very challenging conditions, as further verified in Section~\ref{sec:Results}. 

%%%%%%%%%%%%%%%%%%%%%%%%%%%%%%%%%%%%%%%%%%%%%%%%%%%%%%%%%
\subsection{Bessel Ratio Approximation}\label{subsec:BRapproximation}

A first countermeasure that we introduce in the classical EP algorithm concerns the way in which the Bessel Ratio~(BR)~$I_1(x)/I_0(x)$ is inverted.
Since $|z^{EP}_{k,f/b}|$ are calculated after the EP message~(\ref{eq:pd_EP}), which is in turn computed by applying the moment matching~(\ref{eq:MomentMatch}), we verified that the occasional collapse of their value is largely due to a numerical instability entailed in the approximation of the BR in~(\ref{eq:MomentMatch}) whose inverse, not available in closed form, is necessary to obtain $|z^{EP}_{k}|$. 
In fact, despite the approximation 
\begin{equation}\label{eq:BRapprox_exp}
    I_1(|z|)/I_0(|z|)\simeq \exp (-0.5/|z|) \, ,
\end{equation}
which is commonly adopted in the literature~\cite{ColavolpeModenini2013, SzczecinskiBouaziziAhikam2020arXiv}, is very close to the exact function when $|z|\gg 0$, {it is not equally accurate for lower values of $|z|$, e.g, in the range $[0,5]$ as shown in Fig.~\ref{fig:BRapprox}.
When this traditional approximation~(\ref{eq:BRapprox_exp}) is adopted, the moment matching equation~(\ref{eq:MomentMatch}) becomes
\begin{equation}\label{eq:MM_EXP}
    \ex^{-0.5/|z^{EP}_k|+j\angle z_k^{EP}} = \sum_{m=0}^{M-1} \overline{\alpha}_{k,MIX}^m \ex^{-0.5/|z_{k,MIX}^m|+j\angle z_{k,MIX}^m},
\end{equation}
where, differently from~\cite{ColavolpeModenini2013}, we adopted BR approximation also in the right-hand side. 
In fact, since the exponential approximation is not perfectly overlapped with the true BR, as seen in Fig.~\ref{fig:BRapprox}, using the same \textit{mapping} function in both sides of the equation yields better results. 
In~\cite{CoCoPiVa23}, we followed a different approach, by replacing the BR straightforwardly with its argument $|z|$, so that the resulting moment matching equation,
\begin{equation}\label{eq:Z_EP_ID}
z_k^{EP}= \sum_{m=0}^{M-1} \overline{\alpha}_k^m  z_k^m \, , 
\end{equation}
gives the Tikhonov parameter for the marginal of $\theta_k$ as a simple linear combination of the mixture components' parameters. 
In~\cite{CoCoPiVa23}, our aim was not to accurately approximate the BR but rather to avoid the compression of the $|z_k^{EP}|$ values that, as stated, seems to be inherent to the native EP message passing, due to the shape of the BR (and of its inverse), and tends to depress the level of confidence in the transmitted messages. 
Although this strategy showed satisfactory performance, in~\cite{CoCoPiVa23}, the resulting algorithm requires an additional monitoring of the precision values $|z_{k,f/b}|$, along the forward and backward propagation of $p_{f/b}(\theta_k )$, in order to identify sudden drops of the precision values, which would lead inevitably to errors in phase detection. 
Such a monitoring is implemented in~\cite{CoCoPiVa23} as a {\em post-processing} of the computed precision values, followed by the rejection of messages whose parameters are critical. 
This allows the algorithm to withstand challenging scenarios, e.g, a gap of $4000$ symbols among preamble and postamble without the aid of turbo iterations. 

In this paper, we follow a different approach focusing on the accuracy of the inverse BR approximation.
In fact, because of the very steep shape of the BR in this range, even approximations that are apparently quite similar may produce very different results.  
We analyzed various kinds of approximations and we found that a good compromise between simplicity and performance, in terms of achievable results, is a piecewise function composed by a parabola, for low values of $|z|$, followed by an exponential profile in the saturating range. Therefore, the sought magnitude of the Tikhonov parameter is $|z| = f^{-1}(BR)$, where $f^{-1}(\cdot)$ denotes the following inverse BR approximation, 
\begin{equation}\label{eq:BR_inv}
\!\!\!\!f^{-1}(BR)\!=\!\left\{\begin{array}{cl}
\!\!\!2.55-3.02\cdot \sqrt{0.71-BR},&\!BR\leq 0.59\\
-0.5/\log(BR)+0.55,& \text{elsewhere}\end{array}\right.\!\!\!.
\end{equation}
In Section~\ref{sec:Results}, simulation results demonstrate the significant advantages related to this approximation which lead to a higher accuracy in the estimation of the Tikhonov precision $|z|$ obtained by solving the moment matching equation.

\begin{figure}
    \centering
    \includegraphics[width=1.05\columnwidth]{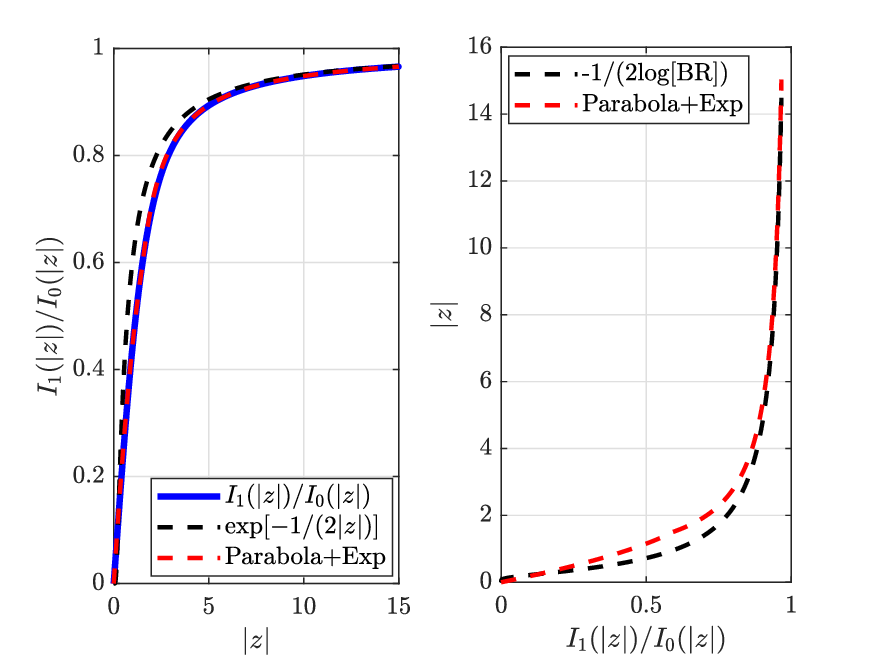}
    \caption{On the left, the comparison among the Bessel ratio, its commonly adopted approximation and the proposed one. On the right, the comparison between the two approximations for the inverse BR.} 
    \label{fig:BRapprox}
\end{figure}

 %%%%%%%%%%%%%%%%%%%%%%%%%%%%%%%%%%%%%%%%%%%%%%%%%%%%%%%%%
\subsection{Rejection Strategy}\label{subsec:Rejection}
 
The rejection of critical messages is, indeed, a well assessed technique within the EP framework, whenever the projection and division operations~(see (\ref{eq:pd_EP})) make {\em improper distributions} arise~\cite{Seeger_PhD_2003, CoPiCoVa23}. 
These distributions are easily recognizable when dealing with Gaussian mixtures, since the resulting message is characterized by a negative variance. 
When, on the contrary, the approximating family is that of Tikhonov distributions, as in the present case, any value for the parameter $z_k$ in~(\ref{eq:Tikhonov}) characterizes a valid distribution. 
Although improper distributions can sometimes be allowed to propagate along the graph, when the EP algorithms faces critical conditions it is usually advantageous to discard the corresponding updates~\cite{CoPiCoVa23arx} and specific rejection criteria~\cite{Seeger_PhD_2003} become useful, and often necessary, for ensuring the success of EP. 

Typically, improper messages arise whenever there is an inconsistency between the two factors of the distribution to be projected in~(\ref{eq:pd_EP}), i.e, the {\em temporary prior}, represented by the parameter $z_{k,u}$ in~(\ref{eq:z_MIX}), and the observation message $p_d(\theta_k)$, which, being a mixture, is represented by the set of values $z^m_k$ in~(\ref{eq:z_MIX}), one for each possible transmitted symbol. 

The {\em modified moment matching equation}, including the approximation discussed in Section~\ref{subsec:BRapproximation}, is
\begin{equation}\label{eq:MMmodified}
    \!\!\!f(|z^{EP}_k|)\ex^{j\angle z_k^{EP}} \!\!= \!\!\sum_{m=0}^{M-1} \overline{\alpha}_{k,MIX}^m  \frac{I_1(|z_{k,MIX}^m|)}{I_0(|z_{k,MIX}^m|)} \ex^{j\angle z_{k,MIX}^m} \, ,
\end{equation}
where the definition of ${\alpha}_{k,MIX}^m$ is analogous to that in~(\ref{eq:alpha_m}) and $\overline{\alpha}_{k,MIX}^m$ are normalized coefficients, as in~(\ref{eq:alpha_norm}) . 
Before the calculation of~(\ref{eq:MMmodified}), a real-time monitoring is introduced on the mixture parameters $z^m_{k,MIX}$ that result from~(\ref{eq:z_MIX}).  
We define the angle 
\begin{equation}\label{eq:gamma_m}
    \gamma^m_k = \arg \left[ z^m_{k,MIX}z^*_{k,u} \right] \, ,
\end{equation}
where the superscript $m$ refers to the $m$-th mixture mode, which quantifies the inconsistency between the phase estimate brought about by each mode and the one provided by the prior distribution. 
Considering the PN evolution until a given time epoch $k$, the argument of $z_{k,u}$ typically yields a stable estimate of phase noise, which should be refined by the current channel observation. 
When this is very noisy, or there is little agreement between the phase angles, i.e., the absolute value of~(\ref{eq:gamma_m}) is large, the inclusion of the observation message can introduce spurious phase slips in the PN estimate. 
We verified that this problem occurs when the estimated marginal distribution of~$\theta_k$ before the EP projection, i.e., $p_d(\theta_k)p_u(\theta_k)$, is a mixture of modes (associated to the complex parameter $z^m_{k,MIX}$) which is spread around the (temporary) prior distribution of phase noise, $p_u(\theta_k)$ (associated to the complex parameter $z_{k,u}$). In this case, the projection produces a rather flat Tikhonov distribution, i.e., characterized by a small precision $\left| z_k^{EP} \right|$ and a phase estimate $\arg \left[ z_k^{EP} \right]$ which gets far from both the prior and the mixture modes, thus creating an uncertainty condition in the estimation process from which it is sometimes difficult to recover, in the absence of pilot symbols. 

In order to avoid such situations, the specific {\em rejection strategy} that we adopt is thus based on the {inconsistency angles}~(\ref{eq:gamma_m}). 
We assume that {the $m$-th mixture mode is inconsistent} if $\left| \gamma^m_k \right| > \Gamma_{th}$, i.e, if its absolute angular deviation from the prior phase angle $\arg \left[ z_{k,u} \right]$ exceeds a given threshold $\Gamma_{th}$. 
Before performing the projection~(\ref{eq:MMmodified}), at each time epoch $k$, the algorithm checks the rejection condition, 
\begin{equation}\label{eq:Rejection}
    \sum^{M-1}_{m=0} \mathds{1} (\left| \gamma^m_k \right| > \Gamma_{th}) > \overline{M} \, ,
\end{equation}
where $\mathds{1}(\cdot)$ is an indicator function equal to $1$ if the condition is true and $0$ otherwise, to verify if the number of inconsistent modes is above a given limit $\overline{M} < M$. 
When~(\ref{eq:Rejection}) is satisfied, the observation message $p_d(\theta_k)$ is considered unreliable and the information contained in it is rejected by setting $z^m_k$ to zero in~(\ref{eq:z_MIX}). 
Hence, both $p_d(\theta_k)$ and $p^{EP}_d(\theta_k)$ are treated as uniform distributions, which do not carry any information on the phase, whose marginal equals that of the prior, i.e., a Tikhonov distribution with parameter to $z_{k,u}$. 

The values of $\Gamma_{th}$ and $\overline{M}$~(\ref{eq:Rejection}) depend on the modulation format~($M$) and on the severity of phase noise~($\sigma_\Delta$). 
Especially at high signal-to-noise ratios~(SNRs), there can be more than one suitable pair and, possibly, multiple rejection conditions can be adopted by logical disjunction~($\vee$) as, e.g., 
\begin{equation}\label{eq:RC2}
\left( \sum^{M-1}_{m=0}\!\!\mathds{1}(\left| \gamma^m_k \right|>\!\Gamma_{th,1}) > \overline{M}_1\!\right) \vee \left( \sum^{M-1}_{m=0}\!\!\mathds{1}(\left| \gamma^m_k \right| > \!\Gamma_{th,2}) > \overline{M}_2\!\right) 
\end{equation}
in the case of two message rejection conditions, whose parameters should be set such that $\Gamma_{th,1} < \Gamma_{th,2}$ and $\overline{M}_1 > \overline{M}_2$}. 
In any case, as stated, the rejection condition is evaluated before the EP projection operation, thus reducing the number of unnecessary operations.

 %%%%%%%%%%%%%%%%%%%%%%%%%%%%%%%%%%%%%%%%%%%%%%%%%%%%%%%%%
 \subsection{Message Damping}\label{subsec:Damping}

{Another solution to overcome the convergence problems of EP, which is known in the literature, is the use of ``{\em damped}" \mbox{updates~\cite{Gelman_arXiv_2014}, \cite{Seeger_PhD_2003}.}}
In iterative parametric message passing, the damping technique consists in propagating a convex combination of the new parameter value and the old one. 
In our case, instead of computing from~(\ref{eq:pd_EP}) 
\begin{equation}\label{eq:no_damp}
{z}^{EP}_{k,d} = z^{EP}_k - z_{k,u}\,, 
\end{equation}
which is the parameter of $p_d^{EP}(\theta_k)$ at the current $n$-th iteration, the linear combination 
\begin{equation}\label{eq:damp}
z^{(n)}_{k,d} = \xi\cdot(z^{EP}_k - z_{k,u})+(1-\xi)\cdot z^{(n-1)}_{k,d} 
\end{equation}
is used, which partly recovers the value computed at the previous, $(n-1)$-th, inner detector iteration. 
The value for the damping parameter $\xi$ must be selected, as discussed in Section~\ref{sec:Results}, where we apply damping and demonstrate its advantages through numerical results.
 
%%%%%%%%%%%%%%%%%%%%%%%%%%%%%%%%%%%%%%%%%%%%%%%%%%%%%%%%%
\section{Message Scheduling}\label{sec:sched}

Regarding message scheduling, for any algorithm, we take advantage of the structure of the FG, which evidences the logical separation of the decoding part from the phase detection part, corresponding to the upper and lower halves of Fig.~\ref{fig:FG_CBC}. 
Indeed,  $P_d(c_k)$ is the extrinsic information on the $k$-th code symbol, sent from the decoder to the phase detector, as well as $p_u(\theta_k)$ which, as stated, represents the prior message in~(\ref{eq:def_pu}) and is thus seen as a temporary estimate of phase noise, sent towards the decoder. 

We shall assume that the phase-detector subgraph operates first, by exchanging {\em horizontal} messages along the Markov chain in the lower part of Fig.~\ref{fig:FG_CBC}, so as to update $p_f(\theta_k)$ and $p_b(\theta_k)$ by message passing (see~\cite{CBC_JSAC_2005}, to recall the rules of computation). 
This horizontal message passing could be iterated, with multiple passes, before sending vertical messages to the decoder subgraph so that the phase estimate could be refined. 
A wide variety of message scheduling can be adopted, such as a \textit{flooding} schedule~\cite{KsFrLo01}, a forward filtering followed by either a backward \textit{filtering} or a backward \textit{smoothing} and so on. 
After a detailed comparison, we concluded that the best performing scheduling is a forward/backward filtering, composed by two independent passes, one for each direction. 
Therefore, during the first inner-detector iteration the temporary prior parameter $z_{k,u}$ is equal to either $z_{k,f}$ or $z_{k,b}$, depending on the considered direction. 
In other words, the opposite message has no impact on the observation message approximation. 
On the other hand, from the $2$nd iteration onwards, the temporary prior is determined by the current message in the analyzed direction entering the variable node $\theta_k$ and the previous iteration's opposite message, i.e.,
\begin{equation}
    z^{(n)}_{k,u} = z^{(n)}_{k,f}+z^{(n-1)}_{k,b}
\end{equation}
during the forward pass and vice versa for the backward pass.
The number of ``inner" detector iterations is referred to as $N_D$.

After $N_D$ phase-detection iterations, {\em vertical messages} $p_u(\theta_k)$ are sent from the phase detector to the upper part of the FG, implementing the decoder, where an inner message passing procedure takes place, until convergence or until a maximum number of decoding iterations is reached. 

The exchange of information, through vertical messages, between detector and decoder can be iterated for $N_{T}$ times,\footnote{In order to indicate the number of inner detector, inner decoder and turbo iterations, the notation $N_D$-$N_C$-$N_T$ is used.} so as to yield a refinement of both the symbols' and the phase samples' estimates, from which the other receiver half can benefit. 
On the other hand, detector-decoder iterations are considered impractical and they are usually avoided in the implementation of digital receivers, especially at high baud rates.

% \textcolor{Green}
Since phase detection is performed before symbol decoding, the forward/backward messages $p_{f/b}(\theta_k)$ propagated at the first {turbo}  iteration cannot benefit from any symbol estimate, hence phase estimation is accomplished (locally) in a non data-aided fashion, while a data-aided phase estimation occurs, instead, at time epochs where pilot symbols are inserted. 
In the case of turbo iterations, on the contrary, a (provisional) soft-decision-directed strategy is implemented, thanks to the extrinsic information $P_d(c_k)$, even for the payload symbols.

%%%%%%%%%%%%%%%%%%%%%%%%%%%%%%%%%%%%%%%%%%%%%%%%%%%%%%%%%
\section{Complexity Analysis}\label{sec:Complexity}

\begin{table*}
\begin{center}
\hspace*{-0.1cm}
\begin{tabular}{|| c || c | c |c||} 
    \hline 
    & \textbf{Two-terms additions} & \textbf{Two-terms multiplications} & \textbf{LUT accesses} \\ [0.5ex] 
    \hline\hline
    \textbf{TP~\cite{CBC_JSAC_2005}} & $7M+12$ & $11M+22$ & $2M+2$   \\ 
    \hline
    \textbf{EP} & $16M+18$ & $34M+25$ & $12M+8$ \\ 
    \hline
    \textbf{dp-BCJR} & $5N_\theta^2+(18M-6)N_\theta-(M+2)$ & $N_\theta(14M+1)+1$ & $2N_\theta^2+N_\theta(3M-1)-M$  \\ 
    \hline
    
    \end{tabular}
    \vspace{0.1cm}
    \caption{\label{Tab:complexity} Complexity comparison per code symbol per iteration}
\end{center}
\end{table*}

We compared the computational complexity of the proposed algorithm based on the EP framework with the TP algorithm proposed in~\cite{CBC_JSAC_2005} and the dp-BCJR.
Referring to the $k$-th time epoch, we considered the operations needed for the computation of the observation message $p_d(\theta_k)$ approximation, the forward $p_f(\theta_k)$ and backward $p_b(\theta_k)$ messages and, finally, the message $P_u(c_k)$. 
In order to compute the algorithmic complexity, we counted the number of required operations distinguishing among two real terms multiplications, additions and accesses to the lookup table~(LUT) for the computation of nonlinear functions.
The total number of required operations is reported in Table~\ref{Tab:complexity}. 
As it can be seen, the complexity of all the algorithms increases linearly with the constellation cardinality~$M$. However, for the dp-BCJR algorithm, it is much higher. In fact,  a critical parameter which has a strong impact on the computational load is the number of discretization levels~$N_\theta$. 
An empirical rule for the choice of this parameter is provided in~\cite{CBC_JSAC_2005}, which suggests that it is linearly related to $M$ according to $N_\theta = 8M$. We found that the value obtained through this rule of thumb is not sufficient for an accurate quantization of the PN variables, which requires at least $512$ levels in the QPSK case.

\begin{figure}
    \centering
    \hspace*{-0.25cm}
    \includegraphics[width=1.1\columnwidth]{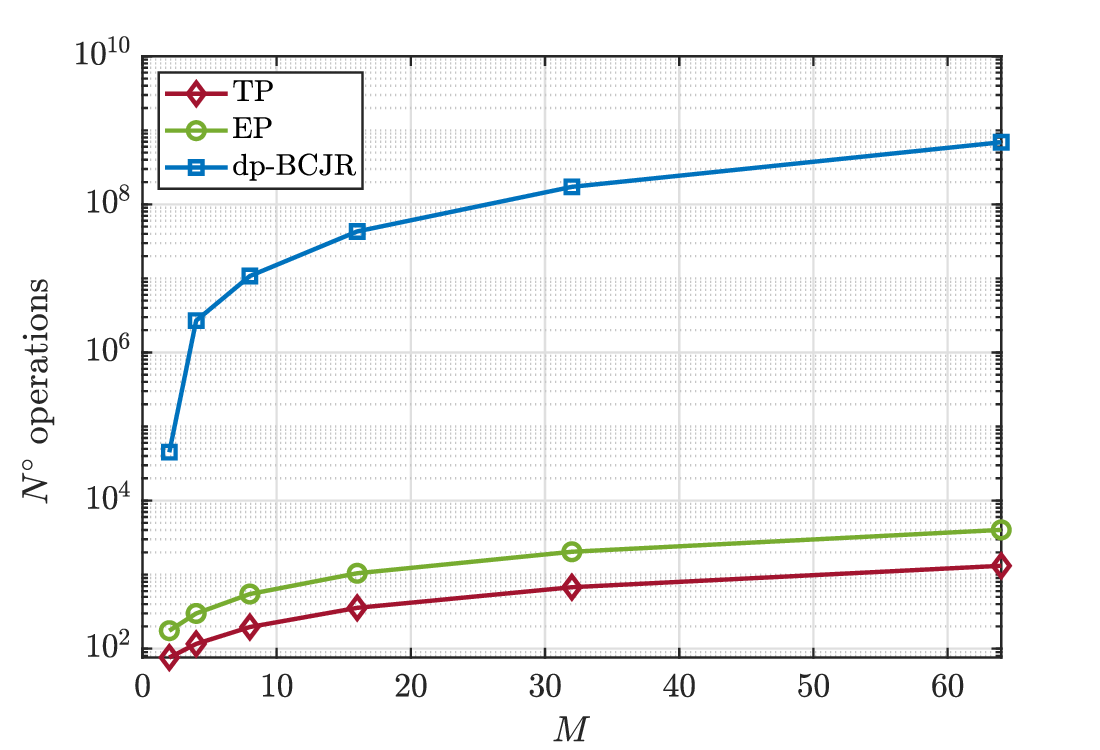}
    \caption{Comparison among the complexity of the analyzed algorithms.}
    \label{fig:Complexity}
\end{figure}

In order to visualize the complexity gap among the analyzed algorithms, the number of required operations is plotted versus $M$ in Fig.~\ref{fig:Complexity}. 
It is possible to notice that the EP and TP algorithms have a comparable complexity whereas the \mbox{dp-BCJR} is characterized by a much higher computational load and hence represents a totally impractical solution. Because of this reason and since, with a sufficiently large $N_\theta$, achieves the performance of the plain SPA, it is considered here just as a performance benchmark. 

%%%%%%%%%%%%%%%%%%%%%%%%%%%%%%%%%%%%%%%%%%%%%%%%%%%%%%%%%
\section{Simulation Results}\label{sec:Results}

We analyzed the performance of different receivers for a coded transmission, employing a low-density-parity-check~(LDPC) encoder, through the system under investigation. Bit error rate~(BER) is plotted versus the SNR~$E_b / N_0$, where $N_0= \sigma^2$ is the variance per component of noise samples in~(\ref{eq:rk}) and $E_b$ is the average energy per information (payload) bit.

In all simulation results, in order to allow the algorithms convergence, pilot symbols, which are known at the receiver, were inserted in the transmitted codeword. The presence of pilots involves a slight decrease of the effective information rate, resulting in an increase in the required average energy per bit $E_b$. In order to make a fair comparison, this increase has been artificially introduced in the curve referring to the {\em Known Phase} scenario, even though pilot symbols were useless in this case since there was no need for phase estimation.
Thus, the {Known Phase} curve shows the performance of the adopted LDPC code.
Furthermore, we included the \textit{All Pilots} curve, which represents an ideal bound since a ``genie-aided'' receiver is assumed to perform the estimation of PN samples resorting to a perfect knowledge of the entire transmitted codeword.
As discussed in Section~\ref{sec:Complexity}, the \textit{dp-BCJR} algorithm is regarded as a performance benchmark. The number of phase discretization levels is set to $N_\theta = 512$ for all the analyzed modulation formats.

\begin{figure}
    \centering
    \vspace{0.03cm}
    \hspace*{-0.5cm}
    \includegraphics[height = 0.76\columnwidth]{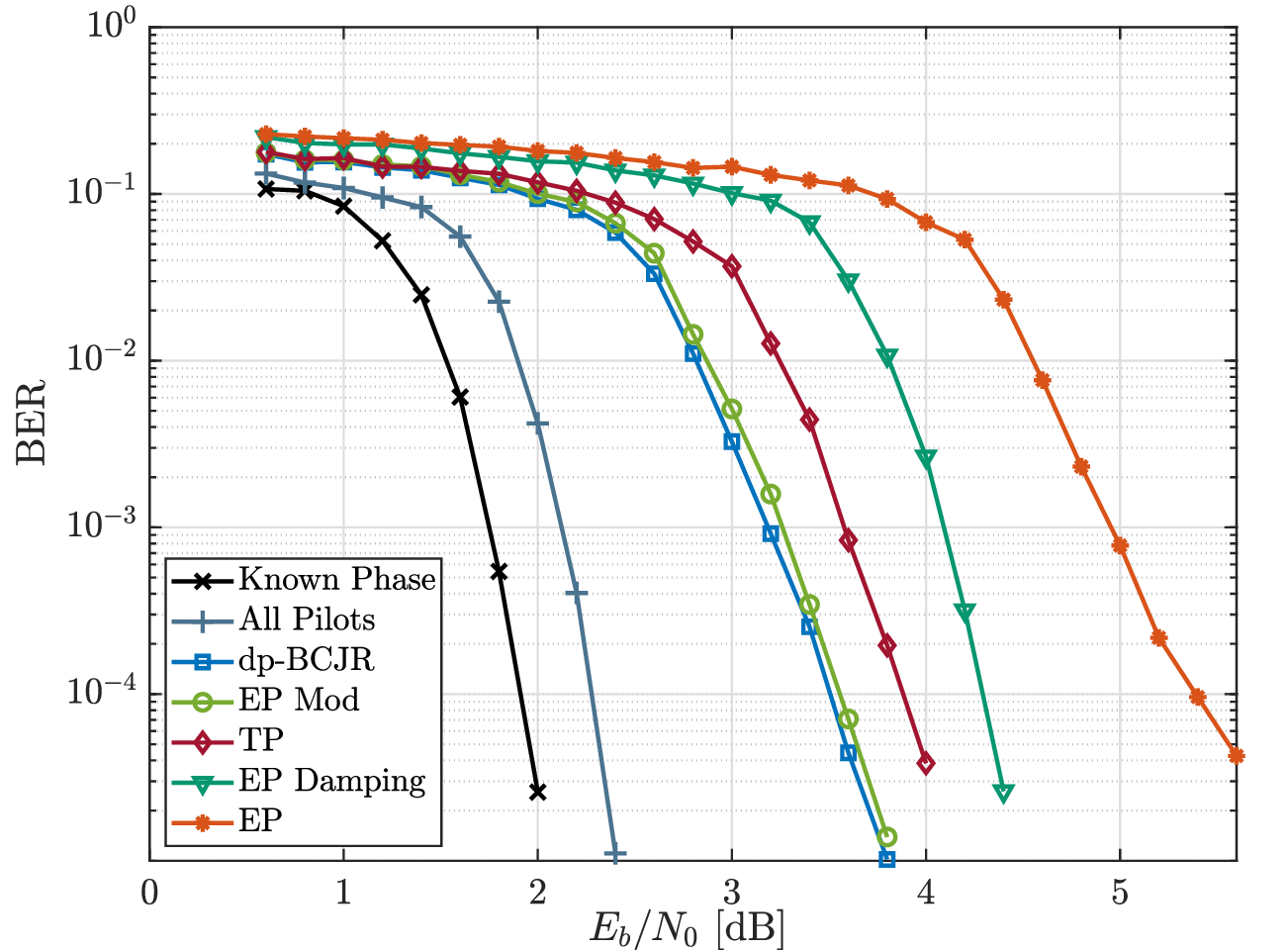}
    \caption{{Performance for an LDPC-coded~\cite{MacKay} QPSK modulation in the separate detection and decoding scenario ($N_{T}=1$). The phase noise standard deviation is \mbox{$\sigma_\Delta = 6\degree$} and $(1/19)$ pilot symbols are distributed among the payload (for other parameters, see text)}.}
    \label{fig:BER_QPSK_McKay_Distributed}
\end{figure}

\begin{figure}
    \centering
    \hspace*{-0.4cm}
    \includegraphics[height = 0.76\columnwidth]{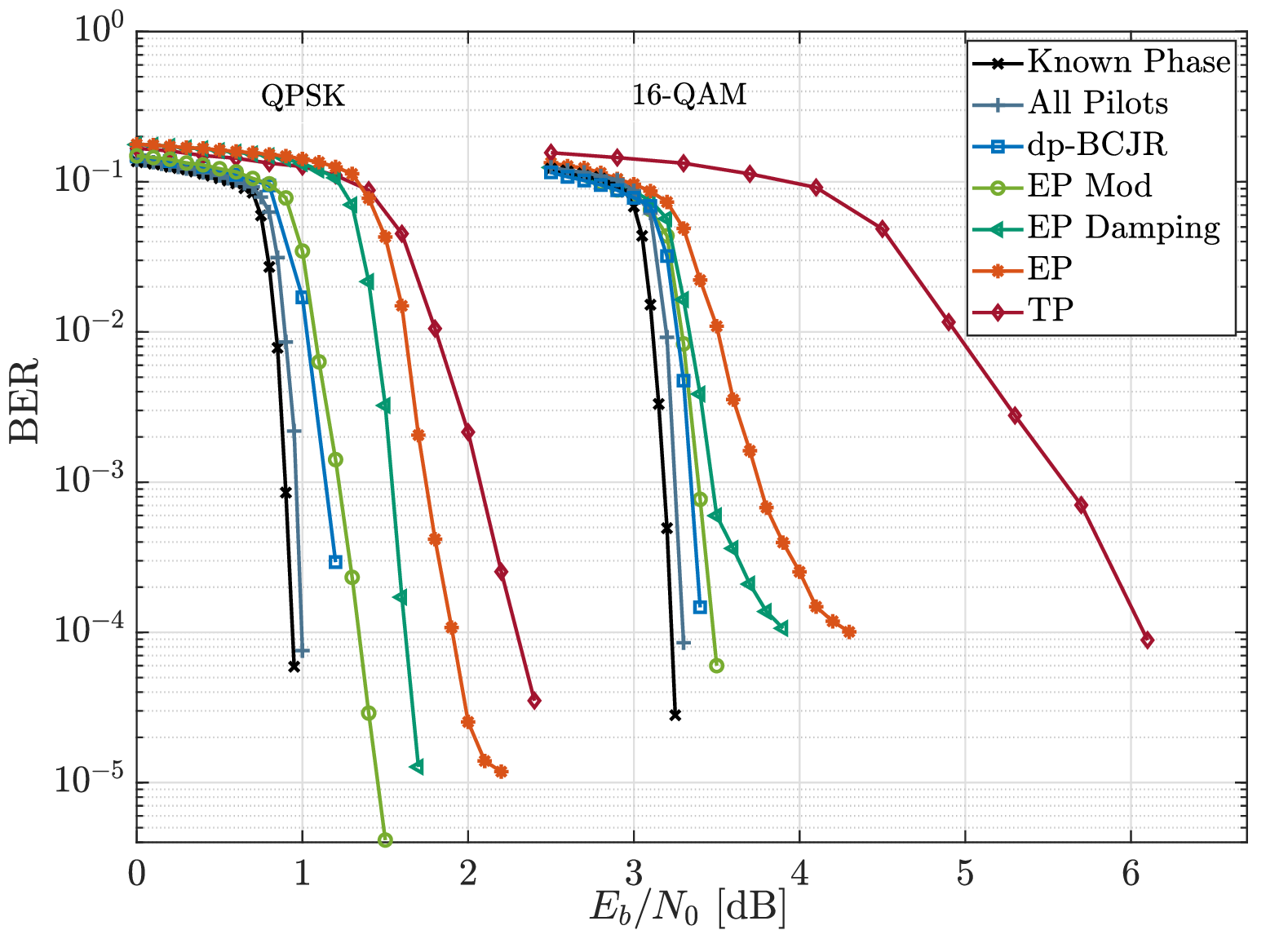}
    \caption{Performance for QPSK and $16$-QAM modulations in the separate detection and decoding scenario ($N_T =1$) employing the DVB-S2 LDPC code (rate $1/2$) and pilots positioning ($36/1440$)~\cite{Et03}. The phase noise standard deviation is $\sigma_\Delta = 1\degree$ (for other parameters, see text).}
    \label{fig:DVB_distributed}
\end{figure}

The performance of the algorithms analyzed in Figs.~\ref{fig:BER_QPSK_McKay_Distributed} and~\ref{fig:DVB_distributed} refers to the practical case of separate detection and decoding, i.e., the absence of turbo iterations ($N_T = 1$). The maximum number of LDPC decoding iterations $N_C$ is set to $200$. 

In Fig.~\ref{fig:BER_QPSK_McKay_Distributed}, we considered the transmission of a block of $2000$ coded symbols punctured by distributed pilot symbols. 
The transmitted sequence started with a pilot, then one pilot symbol was regularly alternated with $19$ payload symbols, amounting to a $5.3\%$ overhead\footnote{We use the notation $\{$number of pilots in a block$\}/\{$payload symbols among two consecutive pilot blocks$\}$ to describe a pilot distribution.}. 
The payload symbols were the output of a  $(3,6)$-regular LDPC code with rate-$1/2$ and length $4000$~\cite{MacKay}. 
The quadrature phase-shift keying (QPSK) modulation format ($M=4$) was adopted, since it is robust to the strong phase noise that was assumed, with $\sigma_\Delta =6^\circ$. 
Although representing an ideal situation, the performance of the All Pilots algorithm shows a loss of about $0.2\,$dB, compared to the Known Phase curve, since PN still has to be estimated in the presence of AWGN. 
The performance of the dp-BCJR algorithm is more than $1\,$ dB away from the theoretical All Pilots reference. 
The TP algorithm of~\cite{CBC_JSAC_2005} shows $0.4\,\text{dB}$ of penalty, compared to the practical dp-BCJR limit. Therefore, thanks to the help of distributed pilots among the payload, a satisfactory performance is achieved by this low-complexity algorithm. 

Moving to the primary object of our investigation, we compared the \textit{native} implementation of the EP algorithm (labelled \textit{EP} in figures), which employs the exponential BR approximation as in~(\ref{eq:MM_EXP}), with two improved variants. 
The first one only exploits the \textit{damping} technique described in Section~\ref{subsec:Damping} and is labelled \textit{EP Damping} in figures. 
The second one relies, in addition to damping, on the inverse BR approximation~(\ref{eq:BR_inv}) and on the message rejection strategies discussed in Section~\ref{subsec:Rejection} (labelled \textit{EP Mod} in figures). 
From Fig.~\ref{fig:BER_QPSK_McKay_Distributed}, we notice that the native EP algorithm loses about $2\,\text{dB}$, compared to the dp-BCJR benchmark, a penalty that increases towards higher SNRs, where a slight change of slope can be noticed.
The introduction of damping allows to partially solve the convergence problems of EP, reducing the loss to $1\,\text{dB}$. 
By numerical simulations, we found that the best value for the damping parameter is $\xi = 0.4$, hence we used this value in all simulations, if not otherwise stated. 
Since the native EP and its damped version do not benefit from inner detector iterations,  
% which instead, highly deteriorate the estimation process. Therefore, 
we set $N_D=1$, so that the application of damping actually reduces to an attenuation of the precision values, i.e., the propagation of message parameter~(\ref{eq:damp}) is nothing but~(\ref{eq:no_damp}) multiplied by $\xi$.

The performance of the Modified EP algorithm (\textit{EP Mod}) shows that the adoption of a more accurate moment matching equation~(\ref{eq:MMmodified}) combined with a  suitable message rejection strategy allows to reach the performance benchmark, with a remarkably lower complexity, compared to dp-BCJR.
Therefore, the Modified EP algorithm outperforms not only the native and damped versions of EP analyzed above but also the simpler TP algorithm. 
In this algorithm, inner detector iterations were indeed useful to refine the phase estimate and $N_D = 2$ was sufficient for achieving the convergence of phase estimation. 
We investigated by simulation the  parameters adopted for the rejection condition~(\ref{eq:Rejection}) and found that $\Gamma_{th} = \pi/2$ and $\overline{M} = 0$ achieve the optimal performance reported in Fig.~\ref{fig:BER_QPSK_McKay_Distributed}. 

In Fig.~\ref{fig:BER_QPSK_McKay_Distributed}, however, the successful operation of the proposed solution occurs in a context where the placement of pilots is quite favorable. 
Nevertheless, in many communications standards the position of pilots is predefined. 
Therefore, with the aim of designing a receiver able to cope with sparser pilots' distributions, in Fig.~\ref{fig:DVB_distributed} we tested the same algorithms analyzed above in the case of the \mbox{DVB-S2} (\textit{Digital Video Broadcasting Satellite Second Generation}) system, where pilot symbols are organized into bursts of $36$ symbols every $1440$ information symbols with the exception of a longer sequence of $90$ symbols in the preamble and postamble~\cite{Et03}. 
We considered the standardized LDPC code with codeword length $64800$ and rate $1/2$ whereas the analyzed modulation formats are QPSK and $16$-QAM. 
For this system, we assumed $\sigma_\Delta = 1\degree$ for the phase noise and we still focus on receivers employing separate detection and decoding, i.e., $N_T =1$.

As seen in Fig.~\ref{fig:DVB_distributed}, the algorithm that is most vulnerable to the sparser distribution of pilots  is TP, as expected. 
In fact, a loss of $1\,\text{dB}$ with respect to the dp-BCJR benchmark is registered for the QPSK modulation while, moving to a higher constellation cardinality, it considerably increases reaching almost $3\,\text{dB}$ for  $16$-QAM.
The penalty is reduced to about $0.7\,\text{dB}$ for both modulation formats, using the native EP algorithm, which however shows the onset of a floor around $10^{-4}$ for the $16$-QAM and around $10^{-5}$ for the QPSK. 
The adoption of damping improves the performance but an error floor is still present for the $16$-QAM while for the QPSK there is still a residual loss with respect to the benchmark. 
It is only with the proposed Modified EP algorithm that the BER curves in Fig.~\ref{fig:DVB_distributed} practically coincide with those of the dp-BCJR benchmark. 
The refinement of phase estimates converges after $N_D = 2$ inner detector iterations. 
The employed rejection condition is in the form~(\ref{eq:RC2}) with $\left(\Gamma_{th,1}, \Gamma_{th,2}\right) = \left(\frac{\pi}{12}, \frac{\pi}{6}\right)$ for both modulations, \mbox{$\left(\overline{M}_1, \overline{M}_2\right) = \left(1, 0\right)$} for the QPSK and \mbox{$\left(\overline{M}_1, \overline{M}_2\right) = (12, 7)$} for the $16$-QAM.

\begin{figure}
    \centering
    \hspace*{-0.1cm}
    \includegraphics[height = 0.76\columnwidth]{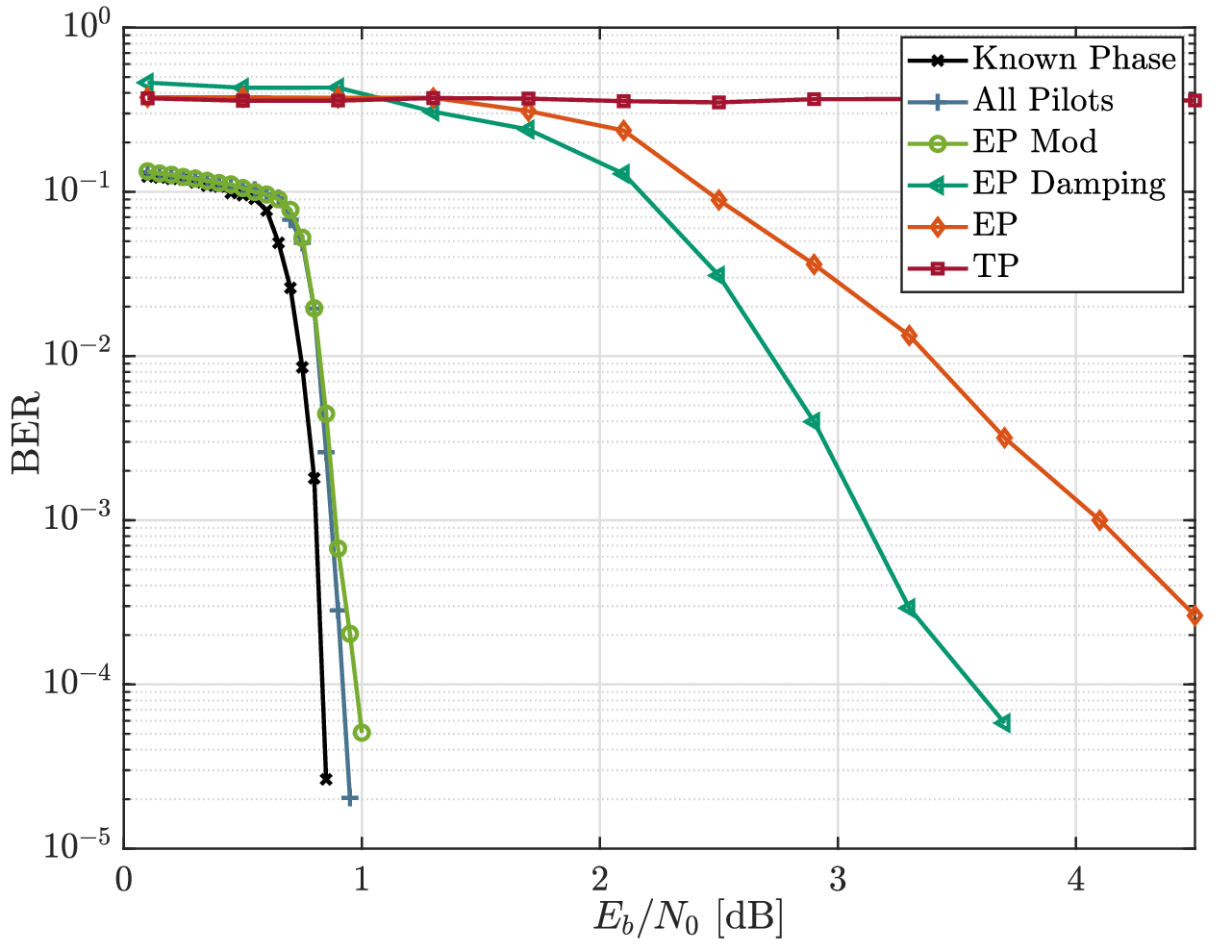}
    \caption{Performance for QPSK modulation in the joint detection and decoding scenario ($1$-$1$-$200$ iterations) employing the DVB-S2 LDPC code (rate $1/2$) and concentrated pilots, positioned in a preamble and postamble each composed by $90$ symbols and separated by $32400$ information symbols. The phase noise standard deviation is $\sigma_\Delta = 1\degree$.
    }
    \label{fig:DVB_concentrated}
\end{figure}

In Fig.~\ref{fig:DVB_concentrated}, we analyzed the most challenging scenario where pilot symbols are concentrated in the preamble and postamble of the data packet, both consisting of $90$ symbols. 
We still employed the DVB-S2 LDPC-code with rate $1/2$ and the QPSK modulation, so that a block of $32400$ information symbols is transmitted without any pilot in-between.
We assumed $\sigma_\Delta = 1\degree$ and, since none of the tested receivers could achieve a satisfying performance when $N_T = 1$, we considered the joint detection and decoding scenario, setting the maximum number of allowed iterations to $1$-$1$-$200$.

Due to the absence of distributed pilots, the TP algorithm fails completely, as expected. 
In fact, in the presence of a (non pilot) payload symbol $c_k$, the observation message $p_d(\theta_k)$ is a uniform pdf, that does not carry any information, since the decoder has no knowledge of the transmitted bits. 
Hence, the TP algorithm cannot bootstrap, even with multiple turbo iterations.
On the contrary, the algorithms based on the EP framework benefit from the exchange of messages between detector and decoder because the PN estimates approximation is carried out after combining the observation message $p_d(\theta_k)$ with the prior distribution $p_u(\theta_k)$. 
Nevertheless, as evidenced by Fig.~\ref{fig:DVB_concentrated}, the native implementation of EP yields unsatisfactory performance and the use of damping allows to gain about $1\,\text{dB}$ but the gap from the ideal {All Pilots} curve is still significant.

The adoption of the proposed improvements leads to an excellent performance achieving an almost perfect overlap to the {All Pilots} curve. 
For this scenario we set $\xi = 0.5$ and we exploited, also inside the rejection strategy,  the knowledge of the current estimate of the symbols pmf $P_d(c_k)$, provided by the decoder. In fact, we set\footnote{We denote with $[\cdot]$ the rounding function to the closest integer.} 
\begin{align}
    &{(\Gamma_{th,1}, \Gamma_{th,2}) = \left(\frac{\pi}{6}, \frac{\pi}{4}\right)\,,}\\
    &{(\overline{M}_1, \overline{M}_2) = ([2\max P_d(c_k)], [\max P_d(c_k)])}
\end{align}
so that the clearer the decoder's belief regarding the transmitted symbol, the more stringent the condition to be met for rejection. Moreover, with reference to the $n$-th turbo iteration, we apply message rejection only if~(\ref{eq:RC2}) and 
\begin{equation}\label{eq:check_Rej}
    {\max_{c_k \in \mathcal{A} } P^{(n)}_d(c_k) \leq \max_{c_k \in \mathcal{A} } P^{(n-1)}_d(c_k)\,,}
\end{equation}
are both satisfied. In fact, condition~(\ref{eq:check_Rej}) identifies the cases where the iterations are not properly refining the symbols and PN estimates. Finally, at high SNRs, the value of $N_0$ that is passed to the detector is slightly larger than the actual one to avoid the overestimation of the message reliability as commonly done in the literature when dealing with suboptimal iterative algorithms. 
%%%%%%%%%%%%%%%%%%%%%%%%%%%%%%%%%%%%%%%%%%%%%%%%%%%%%%%%%
\section{Conclusions}\label{sec:Conclusions}
We analyzed the expectation propagation algorithm operating in a digital receiver for transmissions over channels affected by Wiener phase noise. 
We identified and discussed the weaknesses of the native implementation of EP and examined various supplementary techniques that could be applied to enhance its performance. 
Our investigation highlighted the challenges that arise in applying the EP framework to the considered context and, while solutions taken from the existing literature proved beneficial, they still presented some relevant limitations.

%\ec{
%We demonstrated the successful operation of the EP algorithm for this kind of channels when either the practical separation of phase detection and decoding is required or the absence of distributed pilots is imposed by the adopted standard.} 

By carefully analyzing the message passing procedure, we found that the EP rules of computation for the messages can sometimes induce a collapse of the level of confidence (precision) in the resulting estimates, hence lead to phase slips and subsequent errors.

To address this, we introduced a more accurate approximation of the message projection rule of EP and proposed a novel message rejection strategy aimed at identifying and discarding misleading messages. 
These countermeasures prevent the appearance of spurious phase detection errors so as to reach the performance of the reference benchmark even in the most challenging scenarios where either the separation of phase detection and decoding is required, or the absence of distributed pilots is imposed by the adopted standard.

\bibliographystyle{IEEEtran}
%\bibliography{RefsPhaseNoise2024}

\end{document}